\documentclass{PoS}

\title{Gamma-ray Burst Host Galaxies as Probes of Galaxy Formation and Evolution}

\ShortTitle{GRB Hosts as Probes}

\author{\speaker{Emily M. Levesque}\thanks{Einstein Fellow.}\\
        CASA, Department of Astrophysical and Planetary Sciences, University of Colorado 389-UCB, Boulder,
CO 80309\\
        E-mail: \email{Emily.Levesque@colorado.edu}}

\abstract{Host galaxies are an excellent means of probing the natal environments that generate gamma-ray bursts (GRBs). Surveys of long-duration GRB (LGRB) host environments and their ISM properties have produced intriguing new results with important implications for LGRB progenitor models. These host studies are also critical in evaluating the utility of LGRBs as potential tracers of star formation and metallicity at high redshifts, particularly when considering the implications for properties of host galaxies above z $\sim$ 6. I will summarize our group's latest research on LGRB host galaxies, and discuss the resulting impact on our understanding of these events' progenitors, energetics, afterglow properties, and cosmological applications. }

\FullConference{Gamma-Ray Bursts 2012 Conference -GRB2012,\\
		May 07-11, 2012\\
		Munich, Germany}

\begin{document}

\section{Introduction}
Long-duration gamma-ray bursts (LGRBs), some of the most energetic events observed in our universe, are associated with the core-collapse deaths of young massive stars. As a result of this connection, they are widely cited as powerful and potentially unbiased tracers of the star formation and metallicity history of the universe out to $z \sim 8$ [1,2,3,4]. In recent years, however, potential biases in the star-forming galaxy population sampled by LGRBs have become a matter of debate. Recent work on a small number of nearby LGRBs suggested a connection between LGRBs and low-metallicity environments [5,6]. Nearby host galaxies appeared to fall below the luminosity-metallicity and mass-metallicity relations for star-forming galaxies out to $z \sim 1$ [7,8,9,10]. These results could potentially introduce key biases that would impact the use of LGRBs as effective cosmic probes of galaxy formation and evolution.

A metallicity bias, or some correlation between metallicity and the properties of LGRBs, is predicted by the most commonly-cited progenitor model for LGRBs, the collapsar model [11]. Under the classical assumptions of stellar evolutionary theory, the progenitor is a single rapidly-rotating massive star which maintains a high enough angular momentum over its lifetime to generate an LGRB from core-collapse to an accreting black hole. In addition, LGRBs have been observationally associated with broad-lined Type Ic supernovae [12,13,14,15,16], requiring the progenitors to have shed mass, and therefore angular momentum, as a means of stripping away their outer H and He shells. Mass loss rates for these evolved massive stars are dependent on stellar winds [17], which in turn are dependent on the stars' metallicity [18,19]. For young massive stars, the metallicities of their natal environments can be adopted as the metallicities of the stars themselves. This therefore implies that the wind-driven mass loss rates in high-metallicity environments would rob the stars of too much angular momentum, preventing them from rotating rapidly enough to produce a LGRB and suggesting that LGRBs should either be restricted to low-metallicity environments [20,21,22], or produce weaker explosions at higher metallicities [23]. 

\section{The Role of Metallicity in LGRB Production}
\subsection{The Mass-Metallicity Relation}
We recently conducted a uniform rest-frame optical spectroscopic survey of $z < 1$ LGRB host galaxies, using the Keck telescopes at Mauna Kea Observatory and the Magellan telescopes at Las Campanas Observatory [9,10]. The sample was restricted to confirmed LGRBs with well-associated and observable host galaxies. From these spectra we were able to determine a number of key parameters for the star-forming LGRB host galaxies, including metallicity, ionization parameter, young stellar population age, SFR, and stellar mass. The primary metallicity diagnostic used in this work was the ([OIII] $\lambda$5007 + [OIII] $\lambda$4959 + [OII] $\lambda$3727)/H$\beta$ ($R_{23}$) diagnostic [24,25]; for our full sample we found an average $R_{23}$ metallicity of log(O/H) + 12 = 8.4 $\pm$ 0.3. Our stellar mass estimates were determined using the {\it Le Phare} code [26], fitting multi-band photometry for the host galaxies [4] to stellar population synthesis models adopting a Chabrier IMF, the Bruzual \& Charlot synthetic stellar templates, and the Calzetti extinction law [27,28,29]. The fitting yielded a stellar mass probability distribution for each host galaxy, with the median of the distribution serving as our estimate of the final stellar mass. Our sample has a mean stellar mass of log($M_*/M_{\odot}) = 9.25^{+0.19}_{-0.23}$.

These metallicities and stellar masses were used to construct a mass-metallicity relation for LGRB host galaxies, shown here in Figure 1. For comparison, we compared our results to two samples of star-forming galaxies with comparable redshifts: $\sim53,000$ star-forming SDSS galaxies [30] and 1,330 intermediate-redshift ($0.3 < z < 1$) galaxies from the Deep Extragalactic Evolutionary Probe 2 (DEEP2) survey [31]. Surprisingly, there is a strong and statistically significant correlation between stellar mass and metallicity for LGRB hosts out to $z < 1$ (Pearson's $r = 0.80$, $p=0.001$). Rather than a cut-off metallicity above which LGRBs cannot be formed, the LGRB mass-metallicity relation is offset from the mass-metallicity relation for star-forming galaxies by an average of $-0.42 \pm 0.18$ dex in metallicity. The phenomenological explanation for this is still being explored. It is possible that this offset is directly connected to as-yet-understood physical properties of LGRB progenitors and their natal environments. Alternately, recent results suggest that this may be illustrative of a fundamental relation between stellar mass, star formation rate, and metallicity [32,33,34], although this proposed relation does not fully account for the offset between LGRBs hosts and the general galaxy population [35].
\begin{figure}[h]
\begin{center}
 \includegraphics[width=3.2in]{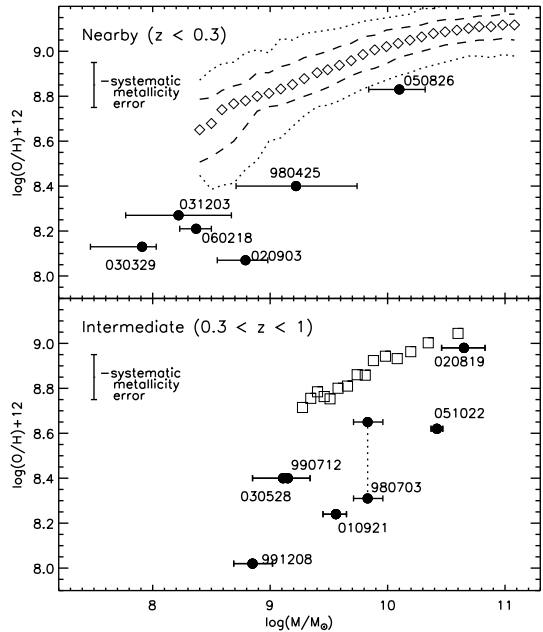} 
 \caption{The mass-metallicity relation for nearby ($z < 0.3$, top) and intermediate-redshift ($0.3 < z < 1$, bottom) LGRB host galaxies (filled circles) [11]. Nearby LGRB hosts are compared to binned mass-metallicity data for the sample of $\sim$53,000 SDSS star-forming galaxies [30], where the open diamonds represent the median of each bin and the dashed/dotted lines show the contours that include 68\%/95\% of the data. For the intermediate-redshift hosts we plot binned mass-metallicity data for a sample of 1330 emission line galaxies from the DEEP2 survey (open squares) [31].}
   \label{fig1}
\end{center}
\end{figure}

\subsection{Spatially-Resolved Host Metallicity Studies} \begin{figure}[h]
\begin{center}
 \includegraphics[width=3.5in]{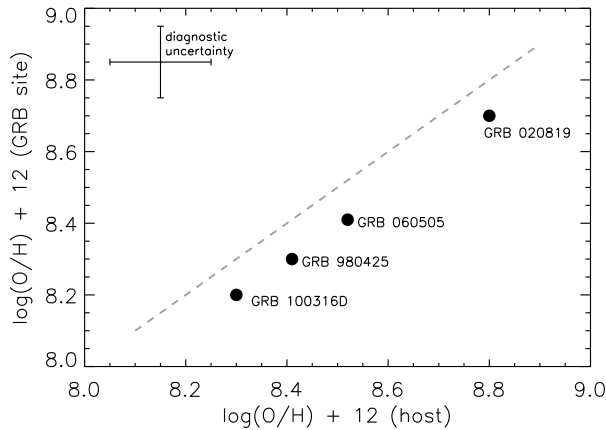} 
 \caption{Explosion site metallicities vs. average host metalicities for the current sample of previously-studied nearby LGRB host galaxies (filled circles) [40]. The theoretical relation where explosion site metallicity and host metallicity are identical is plotted here as a gray dashed line.}
   \label{fig2}
\end{center}
\end{figure} It is important to note one strong limitation of the previous LGRB work, and LGRB studies in general: a reliance on global metallicities. For the majority of LGRB hosts at $z \ge 0.3$, pinpointing the LGRB explosion site and acquiring site-specific spectra within the small, faint host galaxies is a difficult proposition. However, these site-specific studies are possible for a key sample of eight nearby spatially-resolved LGRB host galaxies, including the recent $z=0.28$ host of GRB 120422A [36]. For this subset of hosts we can determine metallicities and star-formation rates directly at the LGRB host site as well as in the surrounding star-forming regions of the galaxy. This allows us to pinpoint the precise environments that produce LGRBs and place these sites in context with their global host properties. Despite the enormous value of such observations, only a small handful of spatially-resolved LGRB hosts have been previously studied. Recent work has obtained integral field unit spectroscopy of the $z = 0.008$ host galaxy of GRB 980425, determining metallicities at 23 different sites across the host [37], and examined spatially-resolved ISM properties in the $z = 0.089$ host galaxy of GRB 060505 [38]. We recently measured high metallicities at both the nucleus and GRB explosion site within the massive $ z = 0.410$ host galaxy of GRB 020819B [39].

Most recently, we have presented a detailed analysis of the GRB 100316D host environment at $z = 0.059$. [40]. By obtaining longslit spectra of the host complex at two different position angles using LDSS3 on Magellan, we were able to extract spatially-resolved profiles for a number of key diagnostic emission features, thus constructing metallicity and star formation rate profiles across the host that focused on both the specific LGRB explosion site and the diffuse emission of the host complex. Based on this analysis, we determined that GRB 100316D happened near the lowest-metallicity and most strong star-forming region of the host complex. However, this work also revealed only a very weak metallicity gradient within the host complex. Combined with the previous studies of nearby LGRB hosts, we found that, within this small sample, ``host" or ``global" metallicities were comparable to metallicities at the GRB explosion sites (Figure 2), suggesting that global metallicities may indeed be valid proxies for explosion site metallicities in higher-redshift LGRB host galaxy studies. Expanding this work to the remaining resolved LGRB hosts (the hosts of GRBs 020903, 030329, and 060218) would allow us to further explore this interesting result.

\subsection{Metallicity and Energetics in LGRBs}
\begin{figure}[h]
\begin{center}
 \includegraphics[width=4.0in]{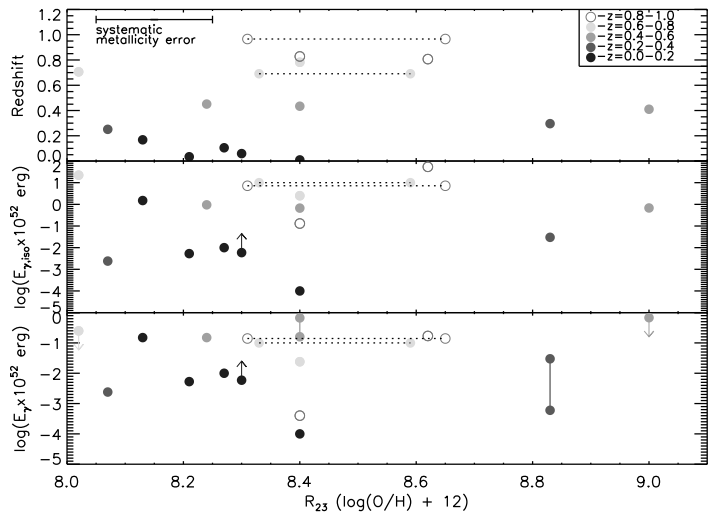} 
 \caption{Metallicity vs. redshift (top), $E_{\gamma, iso}$ (center), and $E_{\gamma}$ (bottom) for our sample of LGRB host galaxies [43]. Hosts are separated into redshift bins to illustrate redshift effects. Upper/lower limits are indicated by arrows; hosts with both limits are shown as data points connected by solid lines.}
   \label{fig3}
\end{center}
\end{figure}
Lacking observational evidence for a pure cut-off metallicity for LGRB formation, we also consider that LGRBs at high metallicity may simply produce less energetic explosions; that is, explosions with a lower isotropic ($E_{\gamma, iso}$) or beaming-corrected ($E_{\gamma} = E_{\gamma, iso} \times 1 - cos(\theta_j)$ energy release in the gamma-ray regime, where $\theta_j$ is the GRB jet opening angle) energy release [41]. Previous studies of several local LGRBs suggested a strong correlation between metallicity and $E_{\gamma, iso}$ [42]. By combining energetic parameters available in the literature with our host metallicities, we were able to reproduce this plot with a larger sample of LGRB hosts (Figure 3) [43]. A comparison with redshift was considered as well. However, we found that there is no statistically significant correlation between metallicity and redshift, {\it or} between metallicity and $E_{\gamma, iso}$ or $E_{\gamma}$. This result is at odds with previous work, and appears to demonstrate that metallicity has no clear impact on the final explosive properties and gamma-ray energy release of an LGRB progenitor.

\section{Looking Ahead}
From the work described above, it is clear that the role of metallicity in LGRB production and progenitor evolution is still a matter of hot debate. These results demonstrate that, while LGRBs occur preferentially in low-metallicity environments, there is no evidence of any low-metallicity cut-off above which LGRB production is suppressed. It also appears that these results cannot be simply explained by metallicity gradients and the presence of low-metallicity progenitors within a globally-sampled host, given that several nearby LGRB host galaxies show evidence of minimal metallicity gradients and progenitor sites that are representative of the global ISM environment. Finally, we have found no statistically significant correlation between the gamma-ray energy release and the metallicity of their host environments. In light of these results, it is worth considering alternative models of LGRBs progenitors and stellar evolution, as well as new analytical means of examining metallicity effects in LGRBs. For example, comparisons with other properties such as X-ray fluence or blastwave velocity could still demonstrate a metallicity correlation.

Alternative progenitor scenarios, such as magnetars or binary channels, also show good agreement with these results. Binary progenitor scenarios in particular are an intriguing possibility. One common binary progenitor scenario for LGRBs invokes a terminal common envelope phase where the outer envelope is ejected and the stellar cores coalesce. This manner of binary is predicted to occur at a higher rate - but {\it not} exclusively - at low metallicity due to weaker stellar winds permitting the evolution of binaries at closer proximities [44]. A second progenitor model considers an interim common envelope phase where the outer envelope is ejected, followed by a contact phase. This scenario should also occur at a higher rate at low metallicity due to a widening range of Roche lobe radii that permit a binary to enter and survive an interim common envelope phase [45].

Finally, in addition to new progenitor scenarios, new treatments of stellar evolution with rotation are also compelling. Detailed treatments of differential rotation have profound effects on the properties and populations of massive stars [46,47], and at low metallicities these effects are expected to be further enhanced [48]. Other recent work has examined the effects of rotation on the production of evolved massive stars, supernovae, and LGRBs, using the new stellar rotation models at solar metallicity [46,49]. This work was able to produce favorable conditions for LGRB formation in 40-60M$_{\odot}$ stars at solar metallicity. Indeed, the latest stellar rotation models actually {\it over}produce the predicted rate of LGRBs, although introducing additional parameters, such as strong coupling of the core and surface due to interior magnetic fields, could decrease this rate and bring predictions into good agreement with observations.

The author is supported by NASA through Einstein Postdoctoral Fellowship grant PF0-110075 awarded by the Chandra X-ray Center, operated by Smithsonian Astrophysical Observatory for NASA under contract NAS8-03060. Collaborators on this work include Edo Berger, Ryan Chornock, Andy Fruchter, John Graham, Lisa Kewley, Alicia Soderberg, and H. Jabran Zahid.

\end{document}